\documentclass[a4paper,oneside,twocolumn,colorlinks=true,linkcolor=blue,citecolor=red]{article}
\usepackage[utf8]{inputenc}
\usepackage[T1]{fontenc}
\usepackage{graphicx}
\usepackage{grffile}
\usepackage{longtable}
\usepackage{wrapfig}
\usepackage{rotating}
\usepackage[normalem]{ulem}
\usepackage{amsmath}
\usepackage{textcomp}
\usepackage{amssymb}
\usepackage{capt-of}
\usepackage{hyperref}
\usepackage{amsmath,amsfonts,amsthm,commath}
\usepackage[version=4]{mhchem} 
\usepackage[mathletters]{ucs}
\usepackage{gensymb}
\usepackage{palatino}
\usepackage{xspace}
\usepackage{romannum}
\usepackage{authblk}
\usepackage[super,sort&compress,comma]{natbib}
\author[1]{Jihoon Shin}
\author[2]{Junghoon Kim}
\author[2,3]{Sung Ha Park}
\author[1,4,*]{Tai Hwan Ha}
\affil[1]{Hazards Monitoring Bionano Research Center, Korea Research Institute of Bioscience and Biotechnology (KRIBB), Daejeon 34141, Korea}
\affil[2]{Department of Physics, Sungkyunkwan University, Suwon 16419, Korea}
\affil[3]{Sungkyunkwan Advanced Institute of Nanotechnology (SAINT), Sungkyunkwan University, Suwon 16419, Korea}
\affil[4]{Department of Nanobiotechnology, KRIBB School of Biotechnology, Korea University of Science and Technology (UST), Daejeon 34113, Korea}
\affil[*]{\textit {email: taihwan@kribb.re.kr}}
\date{}
\title{Kinetic Trans-assembly of DNA Nanostructures}

\begin{document}

\maketitle
\pagenumbering{arabic}

\begin{abstract}
The central dogma of molecular biology is the principal framework for
understanding how nucleic acid information is propagated and used by living
systems to create complex biomolecules. Here, by integrating the structural and
dynamic paradigms of DNA nanotechnology, we present a rationally designed
synthetic platform which functions in an analogous manner to create complex DNA
nanostructures. Starting from one type of DNA nanostructure, DNA strand
displacement circuits were designed to interact and pass along the information
encoded in the initial structure to mediate the self-assembly of a different
type of structure, the final output structure depending on the type of circuit
triggered. Using this concept of a DNA structure `trans-assembling' a different
DNA structure through non-local strand displacement circuitry, four different
schemes were implemented. Specifically, 1D ladder and 2D double-crossover (DX)
lattices were designed to kinetically trigger DNA circuits to activate
polymerization of either ring structures or another type of DX lattice under
enzyme-free, isothermal conditions. In each scheme, the desired multilayer
reaction pathway was activated, among multiple possible pathways, ultimately
leading to the downstream self-assembly of the correct output structure.
\end{abstract}

\section{Introduction}
\label{sec:org200de0a}

The spatiotemporal control of information in any living system is central to
its sustenance. At the cellular level, extra and intracellular information is
carefully processed by a network of biochemical circuitry to orchestrate a
variety of required functions.\cite{lehninger5e} One such process is the
regulation of proteins such as actin and tubulin which themselves polymerize
into higher-order microfilaments and microtubules.\cite{goley06,kim10,brouhard18}
These structures have been studied extensively due to the multiple key roles
they play during the life cycle of the cell, \emph{e.g.} cytokinesis,\cite{goley06}
cell motility,\cite{cameron00} mechanical and structural stability,\cite{revenu04}
intracellular transport,\cite{janke11,akhmanova15} and DNA segregation
\cite{bloom10}. Given the importance of these and other structures in the cell,
it naturally follows that being able to not only synthesize them but also
regulate their synthesis would be critical developments in synthetic biology
and molecular engineering.

To tackle problems in this engineering domain, DNA nanotechnology has provided
some proven solutions,\cite{pfeifer18} namely by way of bottom-up self-assembly
using DNA as a construction material. Progress in this field has largely
occurred on two fronts. The first, and original, motivation is to engineer
structurally intricate yet stable molecular assemblies at the nanoscale using
DNA.\cite{zhang14} This is accomplished by programming strands to bind
specifically through Watson-Crick base pairing to form rigid multiway branches,
which in turn self-assemble into higher-order structures. Prominent examples
from an extensive library include 1D,\cite{liu04,yin08-1,hamada09}
2D,\cite{winfree98,yan03,he05,shin11,wei12} and 3D \cite{zheng09,ke12,ke14,ong17}
tile-based complexes and crystals, and elaborate origami structures
\cite{rothemund06,han11,tikhomirov17,wagenbauer17}. Some of these DNA tiles have
also served as molecular computing elements.\cite{mao00,rothemund04,kim15-2}

The other impetus behind DNA nanotechnology has been the desire to create
dynamic systems, in which DNA assemblies no longer remain static but change
their form, carry out functions, or both. Examples of this type include
autonomous walkers,\cite{bath05,tian05,yin08-2} molecular motors and
robots,\cite{omabegho09,lund10,wickham11,muscat11,douglas12,thubagere17} and
self-replicators \cite{kim15-1}. The central mechanism underlying a large portion
of these works is toehold-mediated strand displacement
reactions,\cite{yurke00,zhang11} where an invading DNA strand latches onto a
short single-stranded toehold domain of a partially double-stranded DNA
molecule and undergoes branch migration, displacing one or more of the
incumbent strands which can further participate in downstream reactions. Strand
displacement reactions have been used to kinetically control DNA self-assembly
pathways which has laid the foundations of a major research theme enabling the
construction of complex DNA circuits such as cascaded digital
\cite{seelig06,qian11-1} and analog \cite{zhang07} DNA circuits, a chemical
reaction network (CRN) to DNA compiler which mimics arbitrary CRN
dynamics,\cite{srinivas17} molecular neural networks,\cite{qian11-2} and molecular
signal amplifiers \cite{chen13}.

Recently, there has been some focus, both theoretical
\cite{schiefer15,schiefer16} and experimental \cite{zhang13,yao15,zhang17}, to
integrate the structural and dynamic paradigms of DNA nanotechnology. These
experimental works typically consist of a two-part scheme in which an upstream
DNA strand displacement circuit interacts with downstream DNA monomers to
facilitate isothermal self-assembly. Here, we advance this concept such that an
initial DNA structure activates a downstream DNA strand displacement circuit
that sequentially leads to polymerization of a new DNA structure, \emph{i.e.}, a DNA
structure transfers information \emph{via} non-local DNA strand displacement
circuitry to kinetically trans-assemble a new DNA structure (initial DNA
structure \(\rightarrow\) DNA strand displacement circuitry \(\rightarrow\) new DNA
structure). We define trans-assembly to be a type of self-assembly in which
information embedded in one structure is released and propagated through
information-relaying media, ultimately triggering the self-assembly of another
structure. This can be conceptually compared to an information-encoded DNA
sequence transferring its information to RNA which mediates the creation of
gene products. Specifically, an initiator strand is introduced into a system
which undergoes toehold-mediated strand displacement with an existing structure
to release a single-stranded signal molecule which triggers one of two DNA
circuits. Depending on which circuit is triggered, another single-stranded
activator molecule is released which binds with a specific type of (inactive)
monomer and initiates strand displacement to render the monomers active. The
activated monomers then polymerize downstream into a new higher-order DNA
nanostructure. Since our system produces only one type of signal strand (per
scheme), the final output structure is limited to periodic structures which,
when compared with \emph{in vivo} RNA and DNA machinery, is rather primitive in
terms of its computational and constructional abilities.

\section{Results}
\label{sec:orga7b4c61}

\subsection{Schematics.}
\label{sec:org5cedb63}

\begin{figure*}
\centering
\includegraphics[width=0.9\linewidth]{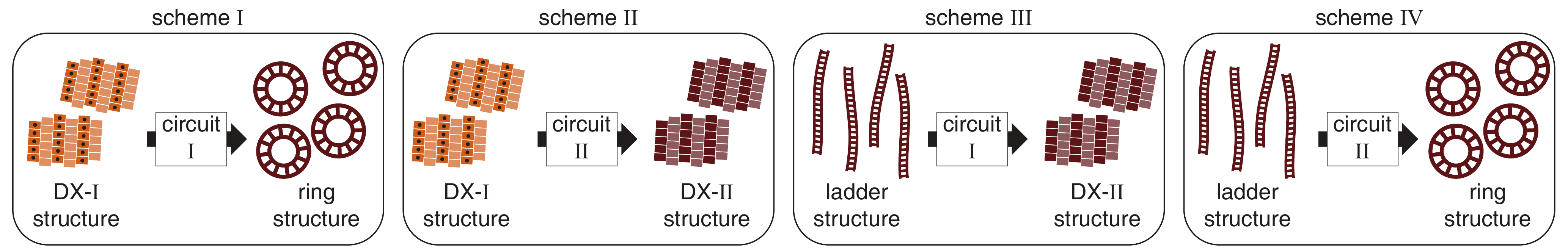}
\caption{\label{fig:fig1}
\footnotesize Kinetic trans-assembly of DNA nanostructures through DNA strand displacement circuitry. The four schemes (\Romannum{1}-\Romannum{4}) implemented in this work.}
\end{figure*}

The schematics of the kinetic trans-assembly of DNA nanostructures we tested in
our work are illustrated in Fig. \ref{fig:fig1} (schemes
\Romannum{1}-\Romannum{4}). All four schemes consist of an initial structure, a
DNA strand displacement circuit, and an output structure. The initial
structures we used in this work were double-crossover (DX) lattices made from
modified DX tiles \cite{winfree98} and ladder structures made from modified
T-motifs \cite{hamada09}. There are two types of strand displacement circuits we
used in this work, circuit \Romannum{1} (used in schemes \Romannum{1} and
\Romannum{3}) and circuit \Romannum{2} (used in schemes \Romannum{2} and
\Romannum{4}). Each of these circuits consists of two subcircuits, subcircuit 1
and 2 (SC1 and SC2) for circuit \Romannum{1} and subcircuit 3 and 4 (SC3 and
SC4) for circuit \Romannum{2}. The final output structure, which depends on
which subcircuit of a given circuit is triggered, is either a discrete ring
structure consisting of another type of modified T-motif or another type of DX
lattice (DX-\Romannum{2}) composed of modified DX tiles.

\begin{figure*}
\centering
\includegraphics[width=1.0\linewidth]{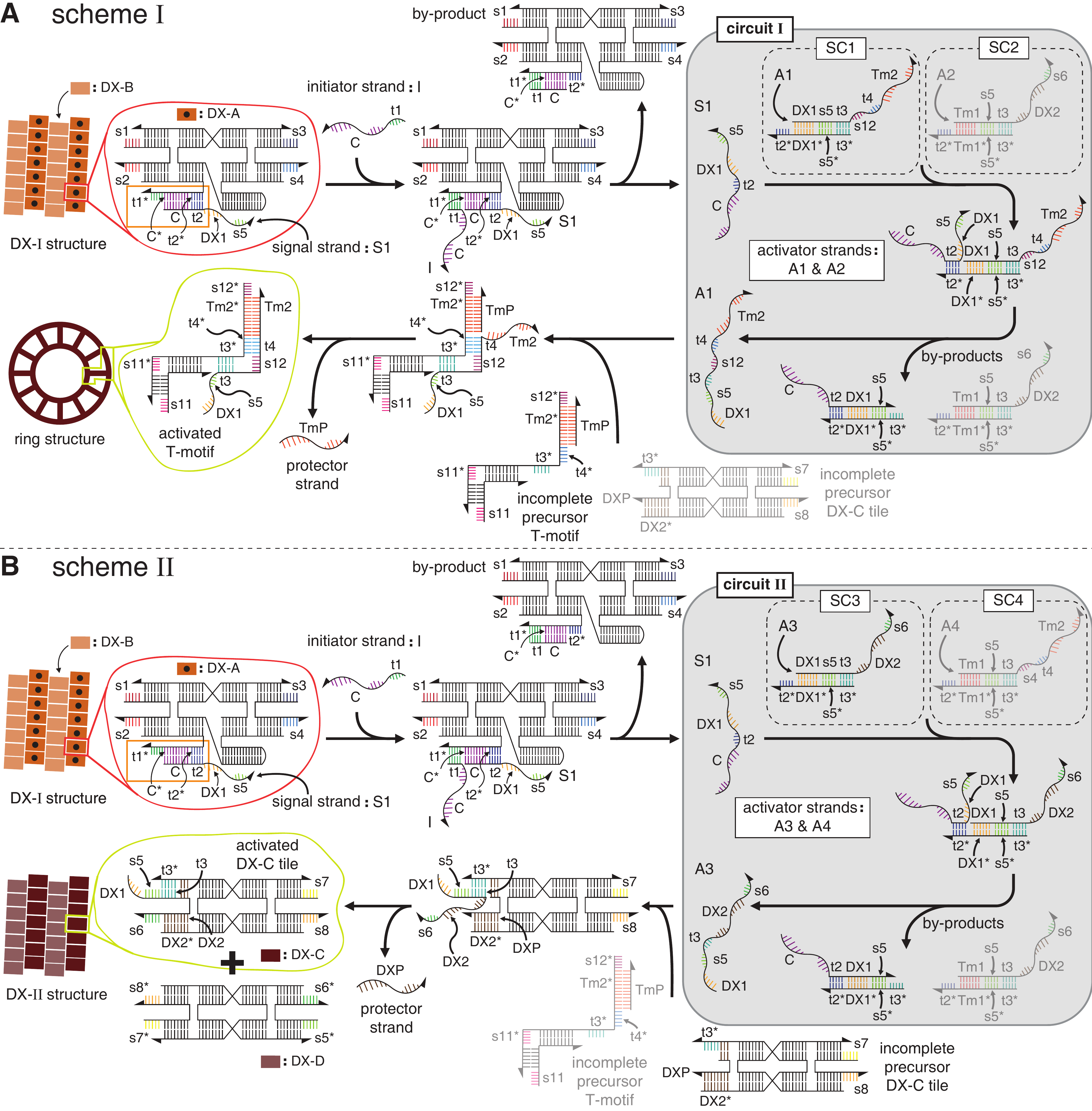}
\caption{\label{fig:fig2}
\footnotesize Schematics of schemes \Romannum{1} and \Romannum{2}. \textbf{a} Details of scheme \Romannum{1}. The reaction starts with an initial DX-\Romannum{1} structure made from DX-A and DX-B tiles. The DX-A tile has an open hairpin region (boxed in orange) which possesses an exposed t1* toehold domain. When the initiator strand I (t1-C) is inserted into the system, its t1 domain hybridizes with the t1* toehold domain of DX-A, which facilitates branch migration of the C domain to displace the signal strand S1 (C-t2-DX1-s5). This signal strand activates circuit \Romannum{1} which includes two subcircuits, SC1 and SC2, who share a common toehold t2* but are otherwise sequentially orthogonal. The t2 domain of the signal strand can bind to the t2* toehold of either SC1 or SC2 but branch migration only occurs for SC1, since it is the only one with a common DX1 branch migration domain. The signal strand for scheme \Romannum{1} activates SC1 of circuit \Romannum{1} which produces the activator strand A1 (DX1-s5-t3-s12-t4-Tm2). This binds to an incomplete precursor T-motif at the t3* and t4* toehold domains. Without the A1 activator strands, these precursor T-motifs cannot self-assemble into higher-order structures due to the missing s12 sticky-end. The domain between the t3 and t4 domains of the invading A1 activator strand provides this missing s12 sticky-end. Once bound at these toehold domains, the A1 strand displaces the TmP protector strand to complete the formation of a new T-motif which self-assembles into a ring structure. \textbf{b} Details of scheme \Romannum{2}. The reaction cascade up to the production of the signal strand S1 is the same as scheme \Romannum{1}. The signal strand for scheme \Romannum{2} activates SC3 of circuit \Romannum{2} which produces the activator strand A3 (DX1-s5-t3-DX2-s6) through strand displacement. After binding at the t3* toehold domain, this A3 activator strand provides the missing s5 and s6 sticky-ends. The s6 sticky-end is provided by displacing the DXP protector strand. This activates the DX-C tile and allows for sticky-end bindings with DX-D tiles to produce DX-\Romannum{2} lattices. All functional domains are colour-coded and labeled; their complementary counterparts are starred (\emph{e.g.} DX1 and DX1*). Half-arrowheads on the strands indicate the 3' direction. Components shown at half opacity stay non-reactive for that particular scheme.}
\end{figure*}

Figure \ref{fig:fig2} depicts the reaction pathways of schemes \Romannum{1} and
\Romannum{2}, respectively. In each scheme, the system initially comprises an
initial structure, two subcircuits, and two types of incomplete precursor
motifs all existing in a single pot. Only the initiator strand, I, is later
added to the system to trigger the reaction cascade. In scheme \Romannum{1},
the initial structure is a DX lattice (denoted DX-\romannum{1}) made up of 2
sequentially different DX tiles, DX-A and DX-B. The DX-A tile of the
DX-\romannum{1} structure is divided into several important functional domains
(Fig. \ref{fig:fig2} a, red delineation). In particular, this tile has one
hairpin protruding from each plane of the tile, one closed and one open. The
open hairpins of the DX-A tiles have a single-stranded toehold domain, t1*
(Fig. \ref{fig:fig2} a, boxed in orange), two consecutive branch migration
domains, C and t2 [henceforth, all consecutive functional domains or strands
will be denoted as their labels conjoined with hyphens going in the 5'
\(\rightarrow\) 3' direction, \emph{e.g.} C-t2, and toehold (sticky-end) domains start
with a lowercase `t' (`s')], and single-stranded DX1-s5 domains. Once an
initiator strand, I (t1-C), is introduced into the system, it binds at the t1*
toehold domain and undergoes strand displacement to displace the incumbent S1
signal strand (C-t2-DX1-s5). The displaced C-t2-DX1-s5 strand acts as a signal
to trigger circuit \Romannum{1}. The two subcircuits of circuit \Romannum{1},
SC1 and SC2, both offer a common single-stranded toehold region t2*, to which
the t2 domain of the signal strand can bind. The double-stranded domains
following the t2* toehold is different for SC1 and SC2 (DX1* and Tm1*,
respectively), meaning branch migration can only proceed along the component
which is complementary to the DX1 domain of the signal strand, \emph{i.e.} the SC1
component of circuit \Romannum{1} is triggered and the A1 activator strand
(DX1-s5-t3-s12-t4-Tm2) is released. Although two types of incomplete precursor
motifs (precursor T-motifs and DX tiles) exist in the system, the released A1
activator strand only interacts with pre-annealed incomplete precursor T-motifs
through the t3 and t4 toehold domains. Without the activator strand, the
precursor T-motif tiles cannot by themselves self-assemble into any specific
higher-order structure due to a missing s12 sticky-end. Instead, this missing
sticky-end is present in one of the domains of the A1 activator strand, and is
provided to the precursor T-motif by way of another strand displacement
reaction. The activator strand has a t3-s12-t4 region where t3 and t4 bind to
their respective toeholds of the precursor tile and the s12 domain remains
single-stranded to fulfill its role as the missing sticky-end. The rest of the
activator strand, \emph{i.e.} domain Tm2, displaces the TmP protector strand from
the precursor tile to add structural integrity and completes the formation of
the T-motif to render it active for self-assembly. The vertical duplexes of the
activated T-motifs (Fig. \ref{fig:fig2} a, yellow delineation) differ in length
where the shorter of the two duplexes has a length of 6 nt and the longer
duplex has a length of 16 nt (both excluding the sticky-ends). This difference
in length leads to an overall curvature so that when these motifs
self-assemble, they form a ring structure as designed.

Starting from the same initial DX lattice structure, a different output
structure can be obtained by changing the DNA circuit (Fig. \ref{fig:fig2} b).
In scheme \Romannum{2}, the reaction profile up to the release of the S1 signal
strand is the same as scheme \Romannum{1}. The circuit used here, namely
circuit \Romannum{2}, has two subcircuits, SC3 and SC4. Again, only one of
these, SC3, is triggered and strand displacement commences along the DX1*
domain. This reaction produces the DX1-s5-t3-DX2-s6 activator strand (A3
strand). In this scheme, the incomplete precursor tiles which interact with the
A3 activator strands are of the DX-type with two missing sticky-ends (s5 and
s6). The A3 activator strand possessing these missing sticky-end domains first
latches onto the t3* toehold region and displaces that DXP protector strand to
complete the DX tile (DX-C tile) and activate it for downstream polymerization
with DX-D tiles to produce DX-\Romannum{2} lattices.

\begin{figure*}
\centering
\includegraphics[width=1.0\linewidth]{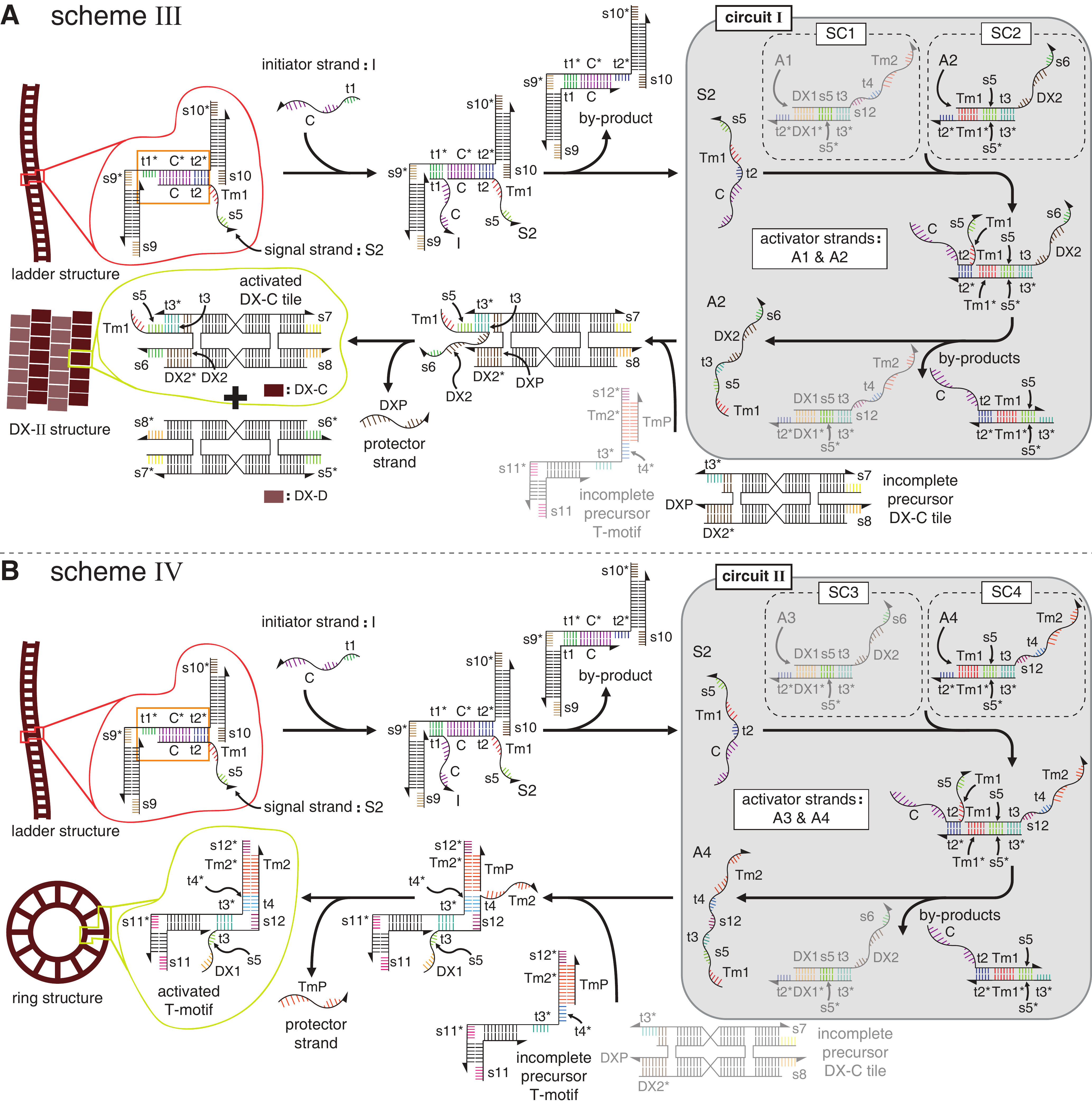}
\caption{\label{fig:fig3}
\footnotesize Schematics of schemes \Romannum{3} and \Romannum{4}. \textbf{a} Details of scheme \Romannum{3}. The initial T-motifs making up the initial ladder structure have an exposed t1* toehold domain (boxed in orange). The reaction cascade starts when the initiator strand I (t1-C) is inserted into the system and its t1 domain hybridizes with the t1* toehold domain of the initial T-motif, which facilitates branch migration of the C domain to displace the signal strand S2 (C-t2-Tm1-s5). This signal strand activates circuit \Romannum{1} which includes two subcircuits, SC1 and SC2, who share a common toehold t2* but are otherwise sequentially orthogonal. The signal strand can bind to the toehold of either SC1 or SC2 but branch migration only occurs for SC2, since it is the only one with a common DX1 branch migration domain. The signal strand then activates SC2 which produces the activator strand A2 (Tm1-s5-t3-DX2-s6) through strand displacement. After binding at the t3* toehold domain, this A2 activator strand provides the missing s5 and s6 sticky-ends. The s6 sticky-end is provided by displacing the DXP protector strand. This activates the DX-C tile and allows for sticky-end bindings with DX-D tiles to produce DX-\Romannum{2} lattices. \textbf{b} Details of scheme \Romannum{4}. The reaction cascade up to the production of the signal strand S2 is the same as scheme \Romannum{3}.  Activation of SC4 produces a displaced activator strand A4 (Tm1-s5-t3-s12-t4-Tm2) which binds to an incomplete precursor T-motif at the t3* and t4* toehold domains. Without the A4 activator strands, these precursor T-motifs cannot self-assemble into higher-order structures due to the missing s12 sticky-end. The domain between the t3 and t4 domains of the invading A4 activator strand provides this missing s12 sticky-end. Once bound at these toehold domains, the A4 strand displaces the TmP protector strand to complete the formation of a new T-motif which self-assembles into a ring structure. All functional domains are colour-coded and labeled; their complementary counterparts are starred (\textit{e.g.} Tm1 and Tm1*). Half-arrowheads on the strands indicate the 3' direction. Components shown at half opacity stay non-reactive for that particular scheme.}
\end{figure*}

The reaction pathways of the other two schemes are shown in Fig. \ref{fig:fig3} a
and b. In these schemes the initial DNA structures are 1D ladder structures and
the output structures are either DX-\Romannum{2} lattices (scheme \Romannum{3})
or ring structures (scheme \Romannum{4}). The reactions proceed similarly to
schemes \Romannum{1} and \Romannum{2} described above. The initiator strand, I
(t1-C), is inserted into the system which displaces the S2 (C-t2-Tm1-s5) signal
strand from the initial structure. This signal strand triggers either the SC2
(SC4) strand displacement subcircuit and produces the A2 (A4) activator strand.
The A2 (A4) activator strand then binds with the incomplete precursor DX-C tile
(T-motif) to activate it for downstream polymerization leading to
DX-\Romannum{2} (ring) structures.

\subsection{Characterization.}
\label{sec:orgf1197c0}

\begin{figure*}
\centering
\includegraphics[width=1.0\linewidth]{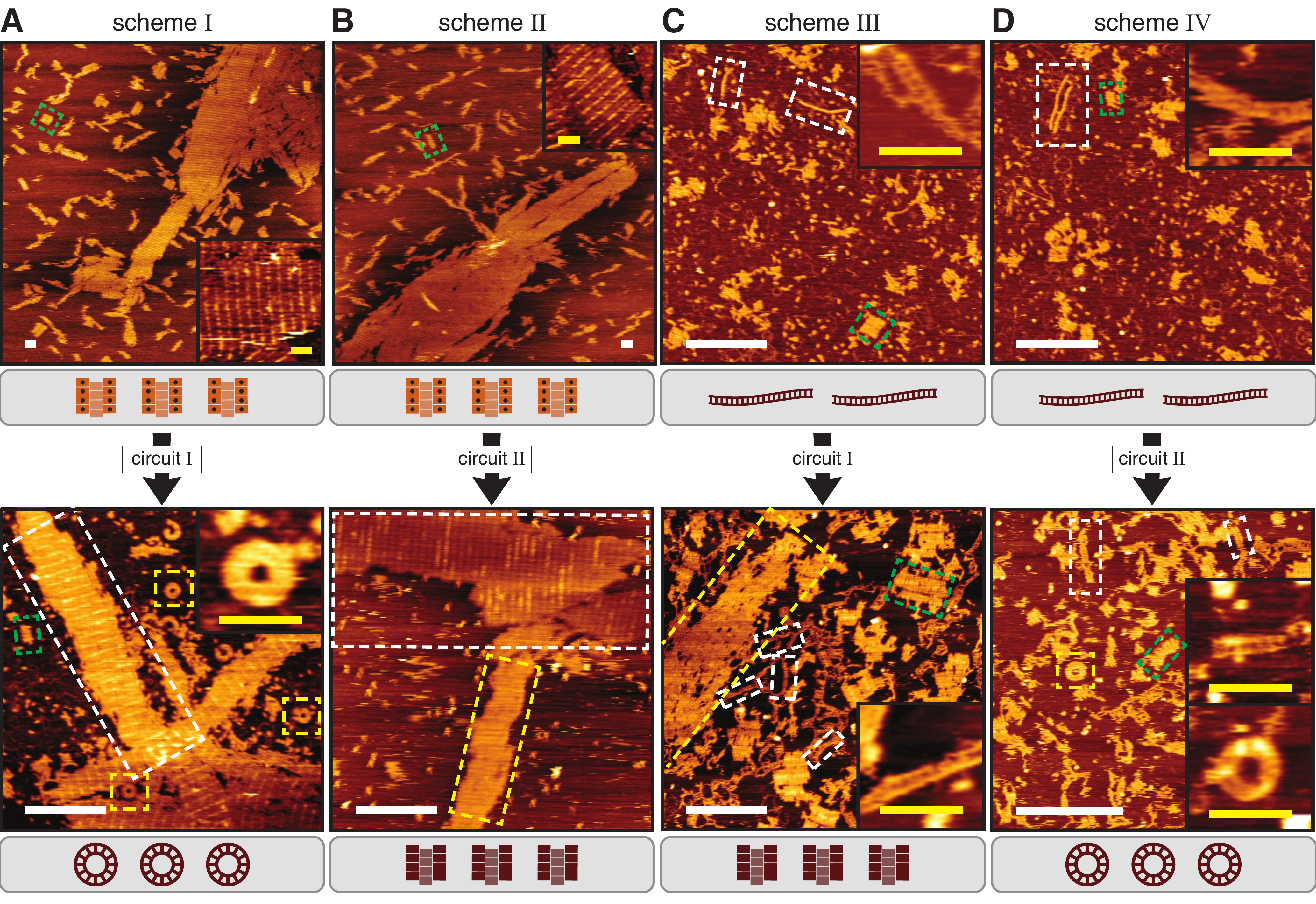}
\caption{\label{fig:fig4}
\footnotesize AFM data of the (top row) initial and (bottom row) final output structures. \textbf{a} Scheme \Romannum{1}. (Top row) The initial DX-\Romannum{1} lattices where the bright stripes are the columns of hairpins. (Bottom row) The self-assembled output ring structures (dashed yellow box) as well as the remaining initial DX-\Romannum{1} lattices (dashed white box). \textbf{b} Scheme \Romannum{2}. (Top row) The same initial DX-\Romannum{1} lattices. (Bottom row) The output DX-\Romannum{2} lattices (dashed yellow box) and remaining DX-\Romannum{1} lattices (dashed white box). The two types of lattices can be distinguished by the presence (DX-\Romannum{1}) or absence (DX-\Romannum{2}) of hairpin stripes. \textbf{c} Scheme \Romannum{3}. (Top row) The initial ladder structures (dashed white boxes) and (bottom row) their output DX-\Romannum{2} structures (dashed yellow box). \textbf{d} Scheme \Romannum{4}. (Top row) The same initial ladder structure (dashed white box) and (bottom row) their output ring structures (dashed yellow box). Some of the images show dashed green boxes which enclose small fragment DX structures formed from bindings between incomplete precursor DX-C tiles and DX-D tiles existing in the system (see Fig. \ref{fig:fig5} d and Supplementary Fig. 8). Insets show magnified views of some of the structures. (White scale bars : 200 nm; inset yellow scale bars : 50 nm)}
\end{figure*}

Characterizations of our schemes were mainly done by atomic force microscopy
(AFM) and agarose gel electrophoresis. Figure \ref{fig:fig4} shows AFM data of
the initial structures and output structures after the strand displacement
circuitry reactions have taken place for each of the four schemes
(Fig. \ref{fig:fig4} a-d, respectively). The DX-\Romannum{1} lattices (top row of
Fig. \ref{fig:fig4} a, b) and initial ladder structures (top row of
Fig. \ref{fig:fig4} c, d) and can be clearly seen. The bright stripes of the
initial DX-\Romannum{1} lattices are columns of hairpins which have a measured
distance of \textasciitilde{}27 nm between each column closely matching the designed distance
of \textasciitilde{}25 nm (Fig. \ref{fig:fig5} a, inset). AFM images of the final output
structures after the systems have gone through the DNA circuitry are shown on
the bottom row of Fig. \ref{fig:fig4}. For the output structure of scheme
\Romannum{1}, discrete ring structures can be easily imaged along with the
remaining DX-\Romannum{1} lattices whose signal strands have been displaced by
the initiator strands (Fig. \ref{fig:fig4} a, bottom row). For schemes
\Romannum{2}-\Romannum{3}, the output structures are hairpin-less
DX-\Romannum{2} lattices (dashed yellow box in the bottom row of
Fig. \ref{fig:fig4} b, c). In scheme \Romannum{4}, initial ladder structures
formed from one type of T-motif are trans-assembled into another type of
T-motif which self-assemble into ring structures (bottom row of
Fig. \ref{fig:fig4} d). AFM analysis of the average sizes of all 4 initial and
output structures is shown in Figure S7.

\section{Discussion}
\label{sec:org272d745}

Some important points concerning our experiments are worth mentioning. 

\subsection{Yield.}
\label{sec:orgb51209e}

One of the determining factors of the final yield of the output structures is
in how well the initial structures form. The initial DX-\Romannum{1} and ladder
structures we designed were assembled from modified DX tiles and T-motifs,
respectively, yet their yields and lattice sizes were comparable to their
original unmodified counterparts.\cite{hamada09,shin11} These initial structures
were then purified by spin filtration before use in our experiments (see
Methods). AFM images of only the annealed DX-\Romannum{1} and ladder lattices
after spin filtration without any other elements or DNA circuitry in the system
show well-formed initial structures (Fig. \ref{fig:fig5} a, b and Supplementary
Fig. 6). Another factor which may be important in the yield of the output
stuctures is the number of reaction layers the information needs to proceed
through in order to produce the final products. This factor may play a role in
reducing the yields of the output structures in our schemes compared with
previously reported schemes with fewer reaction layers
\cite{zhang13,yao15,zhang17}.

\subsection{Prevention of leak reactions.}
\label{sec:orgb9aa3f4}

Another important point to consider when using DNA strand displacement
circuitry is dealing with leak reactions,\cite{zhang07,zhang11} \emph{i.e.} spurious
strand displacement events bypassing the intended reaction pathway and leading
to unwanted signal production. Several preventive measures were taken to
minimize leak reactions. First, the initial structures were purified by spin
filtration before being used in the experiments to remove any single strands
remaining in the system, especially any unattached signal strands (strands
which are not part of the initial structure; see Methods and Supplementary
Fig. 6). All data were obtained using these purified initial structures. As
another preventive measure, the DNA subcircuits and other extant motifs in the
system (SC1-4, incomplete precursor T-motifs and DX-C tiles, and complete DX-D
tiles) were purified before use in order to minimize leak reactions coming from
these elements. Each of the four subcircuits were annealed and run under a 2\%
agarose gel (Fig. \ref{fig:fig5} c). After confirming clear and discrete bands
for all four subcircuits, the bands were extracted and purified using a Freeze
N Squeeze spin column (see Methods). In this way, unhybridized residues were
minimized, especially unhybridized activator strands, while simultaneously
maximizing the use of well-formed ones. This double preventive strategy seems
to be effective since if there were any significant amount of leak reactions,
then observation of the final output structures should be possible without the
introduction of the initiator strand. In other words, at least some output
structures should have been observed in the AFM images before the inclusion of
the initiator strand (Fig. \ref{fig:fig4} top row). The fact that not a single
output structure, either DX-\Romannum{2} lattices or rings, was observed in the
analysis of over 50 images for each scheme before the addition of the initiator
strand strongly suggests the robustness of our DNA strand displacement scheme.
Furthermore, this holds true even in samples which were imaged after \textasciitilde{}7 days of
storage at \ce{4 \degree C}. Only by adding the initiator strand to the system do
the reactions proceed according to design and is the final output structure
produced.

Another potential source of leakage reactions comes from the possibility of the
wrong subcircuit of the DNA circuit being triggered. For example in scheme
\Romannum{1}, only the SC1 subcircuit of circuit \Romannum{1} is designed to be
triggered to release the A1 activator strand but the possibility of the SC2
subcircuit releasing the A2 activator strand (Tm1-s5-t3-DX2-s6) exists. This
may happen either by spontaneous dissociation of the A2 activator strand from
SC2 or effectively due to any single-stranded A2 residues in the system. We do
not believe this to be the case since leakage of A2 activator strands would
provide the missing sticky-ends to the incomplete precursor DX-C tiles and make
them active which in turn would enable bindings with extant DX-D tiles leading
to the self-assembly of DX-\Romannum{2} lattices. As can be seen in
Fig. \ref{fig:fig4} a (bottom image), no DX-\Romannum{2} lattices were observed
(in over 50 AFM images) suggesting negligible production of A2, if any. The
same can be said for schemes \Romannum{2}, \Romannum{3}, and \Romannum{4}; only
the intended activator strand becomes released and insignificant amounts of
leakage occurs for all subcircuits of a given circuit.

\subsection{Control data of the output structures.}
\label{sec:org235aa78}

\begin{figure*}
\centering
\includegraphics[width=1.0\linewidth]{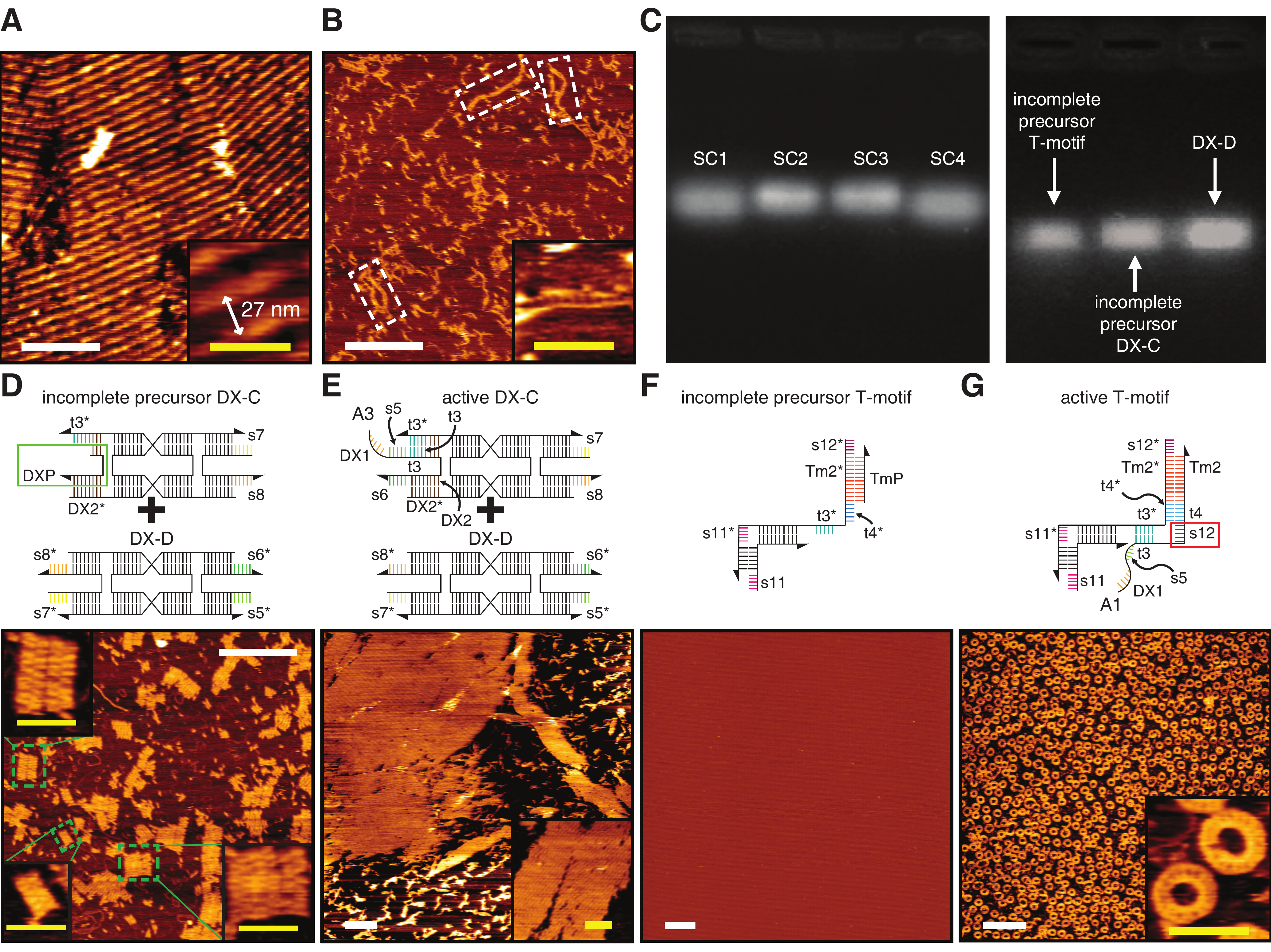}
\caption{\label{fig:fig5}
\footnotesize Control and purification data. AFM data of the initial structures after spin filtration, \textbf{a} DX-I lattices used in schemes \Romannum{1} and \Romannum{2} and \textbf{b} ladder structures used in schemes \Romannum{3} and \Romannum{4}. These two images were obtained prior to adding the other components to the system (\emph{e.g.} DNA circuit components and precursors). The dashed white boxes in \textbf{b} show typical ladder structures. \textbf{c} Agarose gel electrophoresis data of (left) the DNA subcircuits and (right) other motifs used in the experiments. These bands were excised and their products purified and used in schemes \Romannum{1}-\Romannum{4} (see Methods). \textbf{d} Unwanted bindings between incomplete precursor DX-C and DX-D tiles forming small fragment DX lattices. Although precursor DX-C tiles lack s5 and s6 sticky-ends (green box), some bindings between s7, s8, and their complementary sticky-ends occur to form small fragments having widths of (insets) 2-, 4-, and 6-tiles wide (Supplementary Fig. 8). Although it is not too common to find these types of unwanted fragments in schemes \Romannum{1}-\Romannum{4}, this image was taken with a sample made from over \textasciitilde{}13 times the molar concentration of those used in our schemes (400 nM here as opposed to 30 nM samples used in schemes \Romannum{1}-\Romannum{4}) to emphasize this phenomenon. The fragments found in our experiments do not seem to interfere with the intended reaction pathway. \textbf{e} DX-II lattices self-assembled from active DX-C tiles, where the protector DXP strand, green box in \textbf{d}, has been replaced with the A3 activator strand. AFM images of \textbf{f} incomplete precursor T-motifs, which cannot form rings due to missing s12 sticky-ends, and \textbf{g} well-formed ring structures where the A1 activator strand has been added to activate the T-motif by providing the missing s12 sticky-end (red box). Due to the knick in the spoke of the T-motif, the number of motifs that make up each ring structure (\textasciitilde{}20) is a bit higher than conventional T-motif rings (\textasciitilde{}12-16).\cite{hamada09} The samples used in \textbf{d} - \textbf{g} were separately prepared in a one-pot annealing procedure with their respective components at a molar concentration of 400 nM and did not undergo the strand displacement reactions. (White scale bars : 200 nm; inset yellow scale bars : 50 nm)}
\end{figure*}

Figure \ref{fig:fig5} d-g shows control data of the output structures (see
Methods). These structures were individually annealed without any other
elements, \emph{e.g.} DNA circuits, and did not go through the strand displacement
reaction process. Figures \ref{fig:fig5} d and \ref{fig:fig5} e show AFM data of
samples annealed with the two types of tiles making up DX-\Romannum{2}
structures without and with the A3 activator strand, respectively. Without the
A3 strand, only small lattice fragments can be seen (the AFM data shown in
Fig. \ref{fig:fig5} d used samples with concentrations over \(\times 13\) the ones
used in our experiments to emphasize the formation of unintended lattice
fragments, see Supplementary Fig. 8) whereas samples annealed with the A3
strand included show fully-formed DX-\Romannum{2}
lattices. Figure \ref{fig:fig5} f shows that annealing the incomplete precusor
T-motifs by themselves cannot produce rings, but do produce rings when annealed
with the A1 activator strands (Fig. \ref{fig:fig5} g).

\section{Conclusion}
\label{sec:org8c330a4}

Controlled propagation of information emanating from one entity to ultimately
activate or create another entity is ubiquitous in natural and synthetic
biology. Whether it be receptor proteins interacting with a gene regulatory
network for gene expression, membrane receptors binding with primary messengers
to trigger signal transduction cascades leading to cellular responses, or
activated promoters going through synthetic biological circuits leading to a
desired gene product, control of information flow underlies the crucial
metabolic processes of the cell. In this regard, the trans-assembly of an
output structure from an initial structure \emph{via} DNA strand displacement
circuitry we have shown here offers a purely DNA analogue of the aforementioned
biological processes in the context of DNA nanotechnology. The two major
paradigms of DNA nanotechnology, namely the structural self-assembly of DNA
nanostructures and the dynamic transfer of information have been combined into
a single system to provide an isothermal self-assembly platform. Furthermore,
this method can be readily generalized and is not relegated to certain
structures or circuits. Although the experiments in this work were conducted
under a single set of conditions (\emph{e.g.} temperature, salinity, \emph{etc}.), given
the robustness of both the DNA strand displacement reactions
\cite{zhang07,zhang12} and DNA self-assembly, the main results presented here
seem likely to hold for a range of temperatures and conditions. Additionally,
none of the strand sequences for both the motifs and, more importantly, the DNA
circuits were optimized (except checking for repeating segments)\cite{kim16},
providing further support for this assumption.

On a more fundamental level, one may ask whether these types of integrated
systems offer any advantages over implementing DNA tile assembly or DNA strand
displacement circuitry systems separately. It is known that DNA tile assembly
in 2D and 3D is Turing-universal \cite{winfree96} which allows for the
self-assembly of arbitrary shapes from small tile sets \cite{soloveichik07}. As
for CRNs, arbitrary CRN dynamics can be systematically implemented using the
formal language of DNA strand displacement circuitry \cite{srinivas17}. Recent
theoretical works \cite{schiefer15,schiefer16} give some insight into what is
theoretically possible by integrating these two paradigms. For example, the
proposed minimal model in Ref. \citenum{schiefer15} incorporates the non-local
nature of CRNs to influence the local assembly logic of tile assembly, and
\emph{vice versa}. This model is shown to be efficiently Turing-universal even in
(unbounded) 1D. Moreover, arbitrary shapes can be produced using programs
created within this formalism with the program complexity bounded by the
Kolmogorov complexity of the shape, without the sometimes large scale factor
required for programs created within the framework of the abstract tile
assembly model \cite{soloveichik07}. The same authors extend these results in
Ref. \citenum{schiefer16} to show that, in addition to the previously reported
space and program size complexity efficiency,\cite{schiefer15} the efficiency of
computational and construction time complexity of the minimal model is at least
as good if not better than either system alone.

Open questions remain as to what can be fully accomplished by integrating these
two molecular programming paradigms. Perhaps the most far-reaching is, "How
much of the parallel nature of (DNA) molecular computing can these systems
harness?" A new theoretical model or framework may be needed to answer this
question since both of the theoretical works mentioned above enforce sequential
computation/construction.\cite{schiefer15,schiefer16} Even finding physical DNA
tile implementations of the minimal model proposed in Ref. \citenum{schiefer15}
may prove difficult since the abstract tile assembly model is grounded on
different presumptions \cite{winfree98b}. Experimental studies based on a
different type of strand displacement mechanism, namely hybridization chain
reaction, may serve as a guide in answering these questions
\cite{nie14,padilla15,zhang17,helmig17,chatterjee17} but unfortunately a broad
theoretical framework is currently lacking. Although much still needs to be
done, the multilayer, multiple circuit approach of this study to integrate
dynamic information propagation with the self-assembly of DNA nanostructures
shows that the two different DNA chemistries can be carefully coordinated to
mimic the rich and diverse realm of natural biochemical systems.

\section{Materials and Methods}
\label{sec:orgcc53b6d}

\subsection{DNA oligo synthesis}
\label{sec:org97b699c}

Synthetic oligonucleotides were purchased from Bioneer (Daejeon, Korea) and
purified by polyacrylamide gel electrophoresis (PAGE). The details can be found
at \url{www.bioneer.co.kr}. The details of the strand sequences can be found in
the supplementary information.

\subsection{Annealing protocol of complexes}
\label{sec:org0481509}

All the complexes used in schemes \Romannum{1}-\Romannum{4} were self-assembled
by mixing stoichiometric quantities of each strand in a physiological buffer,
1\texttimes{}\ce{TAE/Mg^2+} [Tris-Acetate-EDTA (40 mM Tris, 1 mM EDTA, pH 8.0) with
12.5 mM magnesium acetate]. Each of the complexes [the initial ladder
structures, initial DX-\Romannum{1} lattices, each subcircuit (SC1-4),
incomplete precursor T-motifs, incomplete precursor DX-C tiles, and complete
DX-D tiles] were separately annealed from \ce{95 \degree C} to \ce{25 \degree C}
 by placing the solution-filled microtubes in 1.5 L of boiled water in a
styrofoam box for at least 24 hours to facilitate hybridization. The final
complex/tile/motif concentrations were 2 \ce{\mu M}. For the control samples
used in Fig. \ref{fig:fig5} d-g, each sample was separately prepared in a one-pot
annealing procedure at a tile/motif concentration of 400 nM.

\subsection{Purification of complexes}
\label{sec:org734bff6}

All the complexes obtained by annealing were purified by either agarose gel
electrophoresis or spin filtration. Each of the 2 \ce{\mu M} samples of the
discrete complexes obtained by annealing (\emph{i.e.} SC1-4, incomplete precursor
T-motifs, incomplete precursor DX-C tiles, and complete DX-D tiles) were run
under a 2\% agarose gel and 1\texttimes{}\ce{TAE/Mg^2+} running buffer solution at 60
V for 2 hours. After the run, the gel was stained with SYBR Gold (Thermo Fisher
Scientific, Massachusetts, USA) and imaged with a Molecular Imager Gel Doc XR
system (Bio-Rad, California, USA). To extract gel bands, the gel was placed
under a High Performance 2UV Transilluminator (UVP, Jena, Germany), excised
with a gel cutter, and put into a Freeze N Squeeze microtube (Bio-Rad,
California, USA). The microtube was put in a \ce{-20 \degree C} freezer for 5
min, then removed and immediately centrifuged at 13,000 \texttimes{} r.c.f. (relative
centrifugal force) for 3 min at room temperature. The DNA concentration of
extracted structures was measured by Nanodrop 2000 (Thermo Fisher Scientific,
Massachusetts, USA). The 2 \ce{\mu M} initial (non-discrete) structures, \emph{i.e.}
ladder and DX-\Romannum{1} lattices, were diluted to 800 nM and purified using
a 100kDa molecular weight cut-off (MWCO) Amicon centrifugal filter
(Sigma-Aldrich, St. Louis, Missouri, USA). 100 \ce{\mu L} of the 800 nM samples
and an additional 400 \ce{\mu L} of 1\texttimes{}\ce{TAE/Mg^2+} buffer solution were
added into the filter. After spinning for 3 min at 3,000 \texttimes{} r.c.f., another
400 \ce{\mu L} of 1\texttimes{}\ce{TAE/Mg^2+} buffer was added for a second wash and
repeated three times. The DNA volume/concentration of the filtered initial
structures (ladders and DX-I structures) were typically \textasciitilde{}100 \ce{\mu L}/100 nM
when measured with the Nanodrop 2000.

\subsection{Sample preparation for kinetic trans-assembly}
\label{sec:org8142c39}

A total of four different schemes (schemes \Romannum{1}-\Romannum{4}) were
prepared. For each scheme, stoichiometric amounts of the initial DNA structure,
two DNA subcircuits (SC1 and SC2 or SC3 and SC4), incomplete precursor
T-motifs, incomplete precursor DX tiles, and DX-D tiles were pipetted into a
single microtube. To start the reaction cascade, 10\texttimes{} the amount of the
initiator strand (compared to the other components) was pipetted into the
microtube. The final concentration of the initiator strand was 300 nM and the
final concentration of the rest of the components was 30 nM. The microtubes
were kept at \ce{30 \degree C} for 4 hours to facilitate the strand
displacement and self-assembly reactions. Details of the stoichiometric
quantities of each component used in each scheme can be found in the Supporting
Information.

\subsection{AFM imaging}
\label{sec:org945fece}

To obtain the AFM images, 2 \ce{\mu L} of the samples were placed on freshly
cleaved mica for 30 seconds after which 48 \ce{\mu L} of 1\texttimes{}\ce{TAE/Mg^2+}
buffer was pipetted onto the mica surface. For AFM images of the control
experiments in Fig. \ref{fig:fig5} d-g, 10 \ce{\mu L} of the samples were placed
on freshly cleaved mica for 30 seconds after which 40 \ce{\mu L} of
1\texttimes{}\ce{TAE/Mg^2+} buffer was pipetted onto the mica surface. AFM images
were taken by 3 different instruments. A NanoWizard ULTRASPEED AFM (JPK
Instruments, Berlin, Germany) in Fast imaging mode under a buffer using a wear
resistant high density carbon/diamond like carbon (HDC/DLC) USC-F0.3-k0.3-10
tips (NanoWorld, Neuch\v{a}tel, Switzerland), Digital Instruments Nanoscope
\Romannum{3} (Veeco Inc., New York, USA) in tapping mode under a buffer using
DNP-S10 silicon nitride tips (Bruker, Massachusetts, USA), and MFP-3D-BIO
(Asylum Research, California, USA) in liquid AC mode under a buffer using
DNP-S10 silicon nitride tips (Bruker, Massachusetts, USA).

\section{Acknowledgements}
\label{sec:orgadab73d}
This research was supported by grants from the Korea Research Institute of
Bioscience and Biotechnology (KRIBB) Research Initiative Program and the R\&D
Convergence Program (CAP-14-3-KRISS).

\section{Author Contributions}
\label{sec:org8cb946f}
J.S. and J.K. conceived and directed the project and designed the experiments,
J.S. conducted the experiments, J.S., J.K., T.H.H., and S.H.P. analysed the
data, J.K. and J.S. wrote the paper, and T.H.H. supervised the project. All
authors commented on the manuscript.

\section{Associated Content}
\label{sec:org483c222}
The Supporting Information includes the DNA sequence details and additional AFM
images.

\label{bibliography link}
\bibliographystyle{unsrtnat}
\bibliography{kinetic_DNA}

\pagenumbering{gobble}

\end{document}